\documentclass[10pt, conference, letterpaper]{IEEEtran}

\makeatletter

\usepackage{booktabs} % For formal tables

\usepackage{graphicx}
\usepackage{bm}
\usepackage{amsmath}
\usepackage{subfigure}
\usepackage[ruled,commentsnumbered]{algorithm2e}

\newtheorem{lemma}{Lemma}
\newtheorem{theorem}{Theorem}

\begin{document}
\title{Dynamic Policies for Cooperative Networked Systems}

%\author{George Iosifidis, and Leandros Tassiulas}
%\author{George Iosifidis\\ Trinity College Dublin, Ireland}
%\author{Leandros Tassiulas}
\author{
	\IEEEauthorblockN{George Iosifidis}
	\IEEEauthorblockA{Trinity College Dublin, Ireland}
	\and   %<------ Line breaks in the current column
	\IEEEauthorblockN{Leandros Tassiulas}
	\IEEEauthorblockA{Yale University, USA}
}

\maketitle

\begin{abstract}
A set of economic entities embedded in a network graph collaborate by opportunistically exchanging their resources to satisfy their dynamically generated needs. Under what conditions their collaboration leads to a sustainable economy? Which online policy can ensure a feasible resource exchange point will be attained, and what information is needed to implement it? Furthermore, assuming there are different resources and the entities have diverse production capabilities, which production policy each entity should employ in order to maximize the economy's sustainability? Importantly, can we design such policies that are also incentive compatible even when there is no a priori information about the entities' needs? We introduce a dynamic production scheduling and resource exchange model to capture this fundamental problem and provide answers to the above questions. Applications range from infrastructure sharing, trade and organization management, to social networks and sharing economy services.
\end{abstract}

%\keywords{Economic Networks, Resource Sharing, Maxweight Policies}

\section{Introduction}

Economic entities (EE), being individuals, organizations or countries, need resources (natural resources, services, etc.) in order to sustain their existence and normal activity. Their needs are expressed in terms of requests of a certain amount of a given resource (demands) that are generated at time instances and that should be satisfied either by immediate provisioning of the resource if there is in stock with the entity, or provisioning when will become available to the entity in the near future. Each economic entity has the capability of generating resources either because it is endowed (natural resources) or because of production planning and specialization. Hence each EE generates resources, certain quantities of which become available at certain instances and they either satisfy a pending resource request if there is one.% or \myworries{is stored in an appropriate inventory to satisfy future needs.} 

An economic entity is self sustainable if she can satisfy her own needs for resources in the long run in the sense that for each resource type the rate with which resource is produced exceeds the rate with which resource requests are generated. If resource requests are generated faster than the production rate capability of that resource then there is a shortage of the resource for the entity; this may undermine the long term sustainability of the entity. If we have a collection of EE that are capable of exchanging resources, it is possible that although some of them are not sustainable by their own, they may engage in a resource exchange scheme where an entity covers her own shortage in a resource by the excess production of another EE. When there is an exchange scheme such that all EE become sustainable then the resulting economy is \emph{sustainable}. 

In the first part of this paper we consider an economy of EE specified by an \emph{exchange graph} the topology of which indicates which EE may exchange resource with which other EE, and the rates of resource demands and production of each EE. We specify the conditions for sustainability of such an exchange economy, and we provide a dynamic exchange scheme where each EE determines how to allocate her excess resources to her neighbors such that each EE satisfies her needs if the exchange economy is sustainable. Interestingly enough, no central coordination is necessary and it is adequate if each EE just observes the pending resource requests of her neighbors and allocates the excess resource to the neediest neighbor (including her own needs). 

In the second part of the paper we consider the case where each EE may do some planning of her production capabilities. It is her choice to increase the production of a certain resource by committing more effort to that purpose to the expense of reducing the production of another resource from the production of which the effort is reallocated. The production choices of an EE are reflected to the production rates of the different resources by the EE. We assume that a production plan is represented by the vector of resource production rates of the different resources under that plan. The possible production plan choices are represented by the set of different production rate vectors that are feasible by the EE. It is reasonable to assume that each EE attempts to find a plan that covers her demands, yet this might not be feasible for all entities. 

Therefore, assuming an exchange economy among the EE we consider the question: \emph{is there a choice of production plan for each EE such that the production vectors result in a sustainable economy?} We provide conditions under which that is feasible, and then we introduce a dynamic scheme for each entity determining her production plan that when operates on top of the exchange policy described earlier we result in a sustainable economy. The production planning may operate in a different (slower) time scale than the exchange scheme and again it is dynamic and agnostic on the global economy picture as it does not require knowledge of the EE capabilities in terms of feasible production rates. Each EE reconfigures her production plan at each time in an attempt to satisfy her own needs and that of her exchange peer EE in the best possible ways based on the declared unsatisfied demands of the past. That dynamic policy has as a result global sustainability.

Finally, we extend this analysis to the important case the production induces significant costs. The cooperation and exchange of resources among the EEs can result in a sustainable economy and also reduce the aggregate induced costs compared to the scenario where each EE operates independently. The question that inevitably arises is under what conditions the entities will cooperate, and in particular how they will agree to split the cost-reduction benefits emerging from their collaboration. Leveraging the Nash bargaining solution, we describe the general properties of such incentive-compatible (IC) cooperative solutions. Moreover, we develop dynamic policies that ensure the sustainable operation of the exchange economy while satisfying the IC criterion. Our solution is agnostic on the actual needs and production rates of the EEs, and ensures their fair (and hence self-enforcing) collaboration even without knowing a priori the benefits of their synergy.

The rest of this paper is organized as follows. In Section \ref{sec:single-commodity} we introduce the dynamic exchange model employed to study these cooperative systems and present a simple distributed algorithm that stabilizes the economy whenever this is possible. Section \ref{sec:production} focuses on the richer model with different types of resources and introduces a dynamic production scheduling policy, amenable to distributed execution, that provably stabilizes the economy. In Section \ref{sec:production-costs} we present a model where the different production plans induce different costs, and devise an incentive-compatible policy that ensures the system's sustainable operation. Finally, Section \ref{sec:conclusions} provides a discussion about related works and concludes our study.

\section{Commodity Sharing} \label{sec:single-commodity}

We consider a set $\mathcal{N}$ of $N=|\mathcal{N}|$ economic entities (EE) who produce a set $\mathcal{K}$ of $K=|\mathcal{K}|$ types of resources (or, commodities) over time. The entities are embedded in a directed connected graph $G=(\mathcal{N}, \mathcal{E})$, where the set $\mathcal{E}$ of edges denotes the possible exchanges which are not necessarily bidirectional. Each EE has two roles, acting both as a \emph{consumer} and as a \emph{producer} of resources. Therefore there are $i=1,2,\ldots,N$ consumers and $j=1,2,\ldots,N$ producers in the system. We denote with $\mathcal{N}_i$ the set of $j=1,\ldots,N_i$ producers who can serve consumer $i$; and with $\mathcal{N}_j$ the set of $i=1,\ldots,N_j$ consumers who can receive resources from producer $j$. The connectivity among consumers and producers is determined by $\mathcal{E}$. Note that $(i,i)\in\mathcal{E}$, $\forall i\in\mathcal{N}$, as each EE can serve her own requests. We first study the setting with one commodity. 

\begin{figure}[t]
	\centering
	\subfigure[]{
		\includegraphics[width=0.16\textwidth]{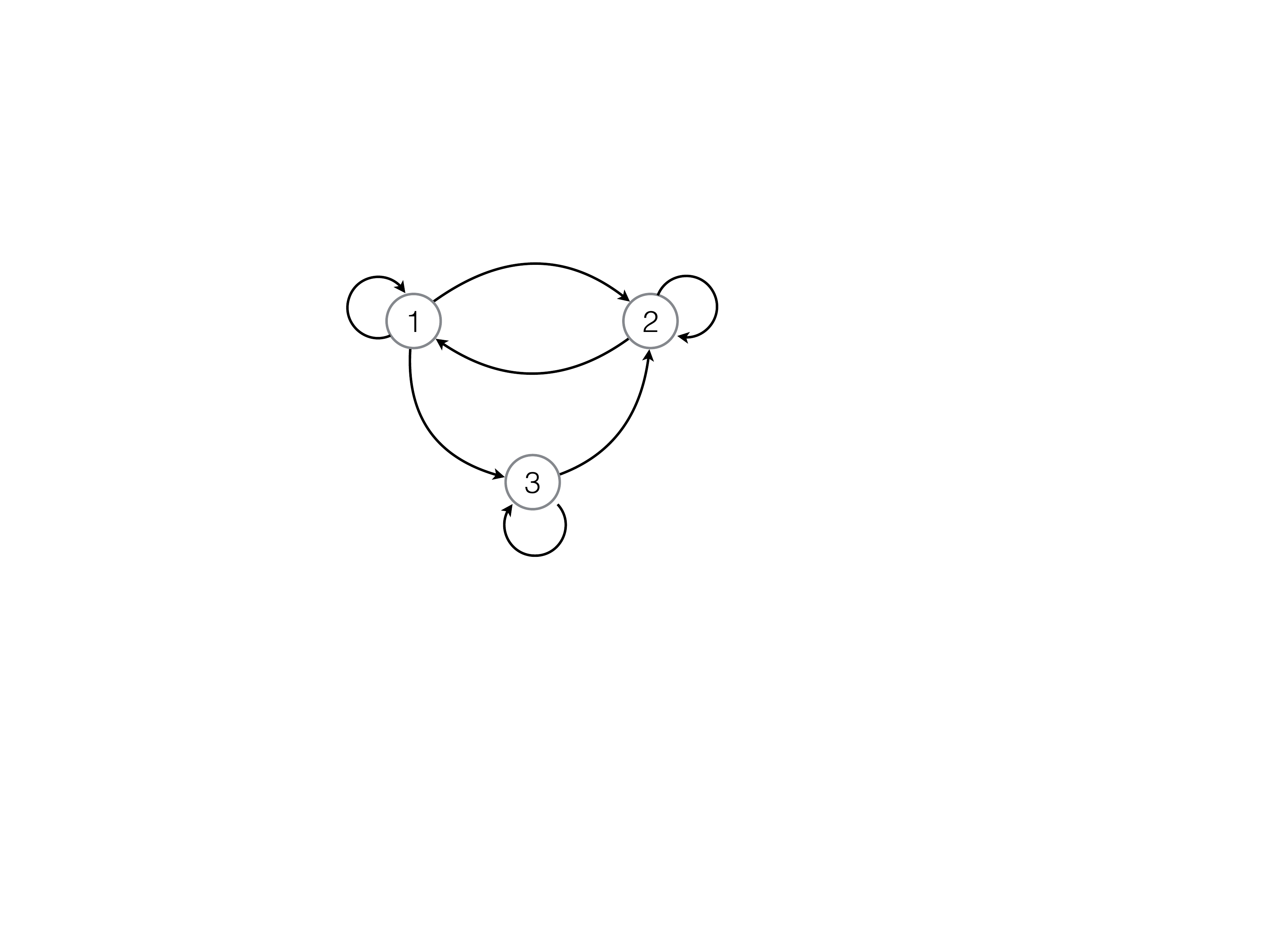}
	}
	\subfigure[]{
		\includegraphics[width=0.21\textwidth]{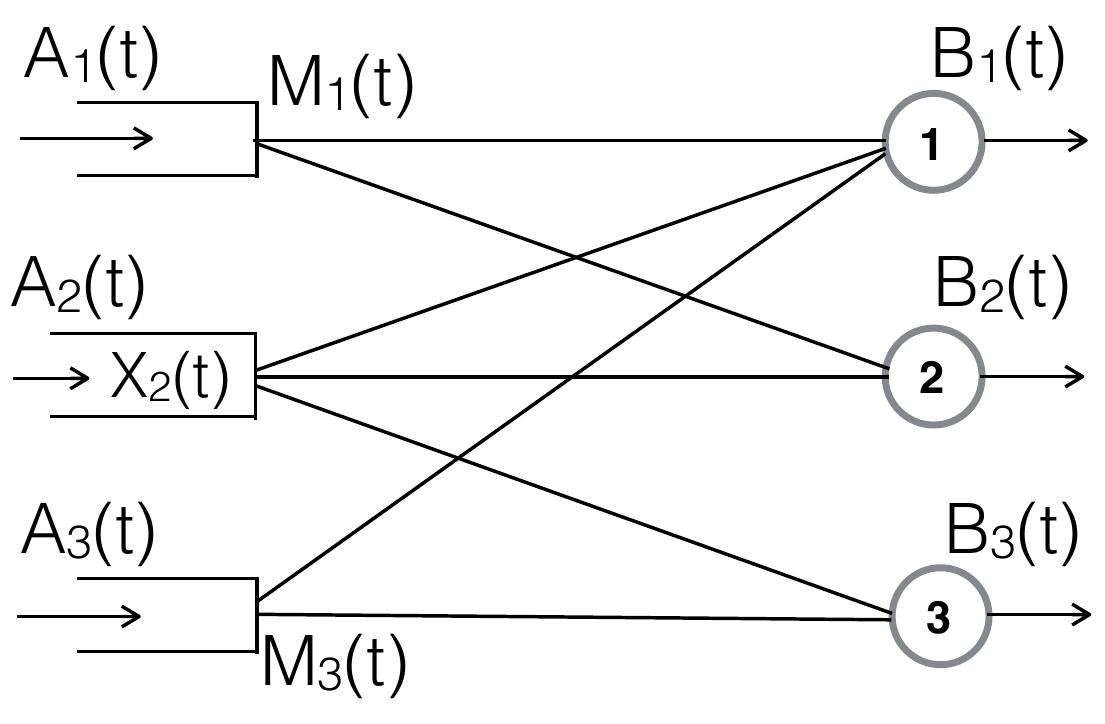}
	}
	\caption{\small{An instance of a cooperative economy with 3 entities and the respective dynamic exchange model with 3 consumers and 3 producers. Each entity can serve its own demands, and the demands of its neighbors. The graph is directed.} }\label{fig:queue-example}
\end{figure}

We assume that resources are produced and allocated in batches, hence we consider a time slotted operation. During each slot $t$, a number of $A_i(t)\geq 0$ resource requests are generated at consumer $i\in\mathcal{N}$. Let $X_i(t)$ be the number of demands, i.e., pending requests, at the $i$th consumer by the end of slot $t$. In the beginning of each slot $t$, $B_j(t)$ units of resource are generated at producer $j$. We define the vector $\bm{B}(t)=(B_j(t),\,j\in\mathcal{N})$. The processes $A_i$, $B_j$, $i,j\in\mathcal{N}$, are independent and i.i.d. over time, with constant and deterministically bounded expectations  $E[A_i(t)]=a_i\leq A_{max}$, and $E[B_j(t)]=b_j\leq B_{max}$. An example is shown in Fig. \ref{fig:queue-example}.  

The control action in this system is to decide how the resources of the producers will be allocated to the consumers at each slot $t$. Let $I_{ji}(t)\in\{0,1\}$ denote whether producer $j$ is servicing the demand of consumer $i\in\mathcal{N}_j$ during slot $t$, by allocating to $i$ all its available resources. The control matrix is then:
\begin{align}
	\bm{I}(t)=\big( I_{ji}(t)\in\{0,1\}\,:\,(j,i)\in\mathcal{E}\big)
\end{align}
for modeling purposes, we assume that $\bm{I}$ is a $N\times N$ binary matrix, where the entry $(j,i)$ can be equal to $1$ only if $(j,i)\in\mathcal{E}$. Without loss of generality, we assume that each consumer can be served by many producers, but each producer can serve at most one consumer or idle. Hence, the set of eligible control matrices is
\begin{equation}
	\mathcal{I}=\{\bm{I}: I_{ji}\geq 0,\,\, \sum_{i\in\mathcal{N}_j}I_{ji}\leq 1,\,\,\forall\,i,j\in\mathcal{N}\}. \label{eq:control-constraint}
\end{equation}		
Let $M_i(t)$ denote the aggregate resource that consumer $i$ receives in slot $t$:
\begin{equation}
	M_i(t)=\sum_{j\in\mathcal{N}_i}I_{ji}(t)B_{j}(t)\,.
\end{equation}
Then, the number of its pending requests evolves as
\begin{equation}
	X_i(t)=[X_i(t-1)-M_i(t) ]^++A_i(t)\,,
\end{equation}
and we define $\bm{X}(t)=(X_i(t),\,i\in\mathcal{N})$.

The economic sustainability can be defined using the strong stability requirement for the demands \cite{tassiulas-book}:
\begin{equation}
	\lim_{t\rightarrow \infty}sup\frac{1}{t}\sum_{\tau=0}^{t-1}\sum_{i=1}^{N}E\left[X_{i}(\tau)\right]<\infty\,,
\end{equation}		
or, in other words, we ask that the M.C. $\bm{X}=\{\bm{X}(t)\}_{t=1}^{t=\infty}$ is ergodic and possesses a stationary distribution. It can be easily shown that the necessary conditions for sustainability of this economy are:  
\begin{equation}
	\sum_{i\in \mathcal{Q}}a_i\leq \sum_{j\in\mathcal{N}_{\mathcal{Q}}}b_j,\,\,\,\,\forall\,\mathcal{Q}\subseteq\mathcal{N}\,, \label{eq:stability-condition}
\end{equation}
where $\mathcal{N}_{\mathcal{Q}}$ is the set of producers that can serve one or more consumers in set $\mathcal{Q}$. These conditions characterize the \emph{sustainability region} $\bm{\Lambda}$ of the cooperative economy, i.e., the closure of set of demand generation rates $\bm{a}=(a_i:i\in\mathcal{N})$ that can be supported by the $\mathcal{N}$ entities if they collaborate. We refer to such a cooperative economy as \emph{sustainable}. 

The following Lemma explains that (\ref{eq:stability-condition}) is sufficient in the sense that it guarantees the existence of a random and state-independent servicing policy $\Pi_{opt}$ that allocates with a certain probability the resources of each producer to every consumer she is connected to. 
\begin{lemma}
	For any demand generation vector $\bm{a}\in\bm{\Lambda}$, there exists a stationary allocation policy $\Pi_{opt}$ that renders the economy sustainable. Moreover, this policy can be found in polynomial time.
\end{lemma}	
The allocation probabilities can be also interpreted as time shares of service each consumer receives from her neighbors.

\begin{figure}[t]
	\centering
	\subfigure[]{
		\includegraphics[width=0.22\textwidth]{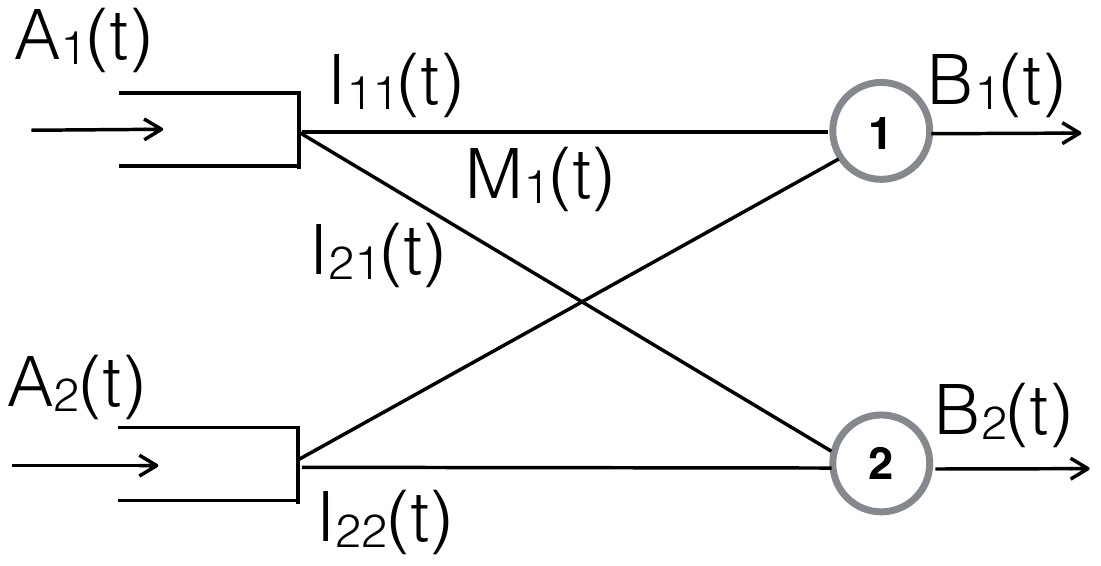}
	}
	\subfigure[]{
		\includegraphics[width=0.22\textwidth]{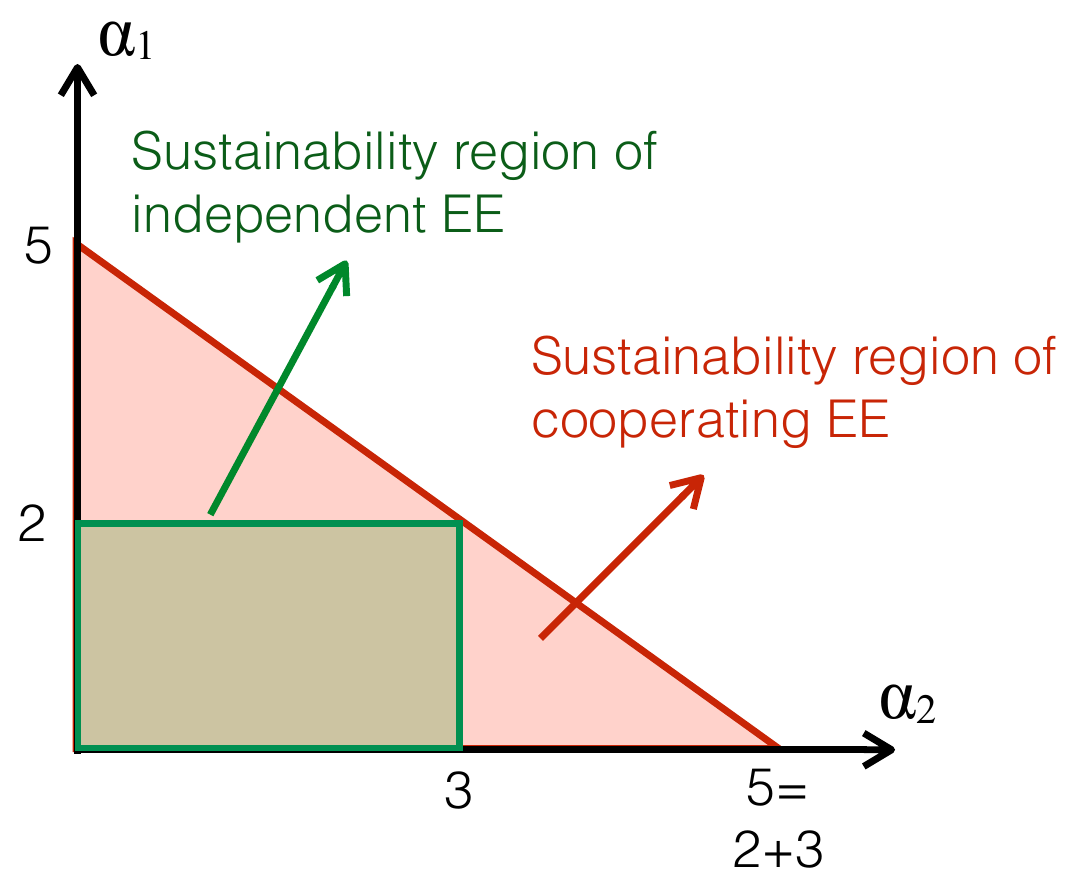}
	}
	\caption{\small{(a): An example of a cooperative economy with 2 EE; it is with $b_1=E[B_1(t)]=2$, $b_2=E[B_2(t)]=3$. (b): Sustainable demand region of the economy when the nodes do not collaborate (green area) and when the nodes do cooperate (red area).} }\label{fig:queue-example-2}
\end{figure}

The next question is whether there exists a dynamic exchange policy that ensures sustainability whenever that is possible. Namely, we are interested in policies that are amenable to distributed implementation and achieve this goal without any prior knowledge about the request and resource generation rates $a_i$, $b_j$, $i,j\in\mathcal{N}$, nor the graph $G$. 

The following theorem describes a policy that satisfies the above requirements and ensures the sustainable operation of the economy whenever $\bm{a}\in\bm{\Lambda}$:
\begin{theorem}
	The max-weight policy \cite{tassiulas-dynamic-server}, that selects in each slot $t$ the control action ${I}(t)\in\mathcal{I}$ such that:		
	\begin{equation}
		I^{*}(t) =\,\,\,\arg\max_{I\in\mathcal{I}} \sum_{i=1}^{N}X_{i}(t)\sum_{j\in\mathcal{N}_i}I_{ji}B_{j}(t)
	\end{equation}
	stabilizes the system if $\bm{a}\in\bm{\Lambda}$ and yields average backlog:
	\begin{equation}
		\lim_{t\rightarrow \infty}\frac{1}{t}\sum_{\tau=0}^{t-1}\sum_{i=1}^{N}E[X_i(\tau)]\leq \frac{NA_{max}^{2}+\sum_{i=1}^{N}d_{i}^{in}B_{max}^{2}}{2\epsilon(\bm{a})} \nonumber
	\end{equation}
	where $\epsilon(\bm{a})$ is the distance of $\bm{a}$ from the boundary of $\bm{\Lambda}$, and $d_{i}^{in}$ the in-degree of EE $i$ in $G$.
\end{theorem}
Due to the design assumption that a consumer can be served by more than one producer the optimal $I^{*}(t)$ can be found in a \textbf{distributed fashion} as shown in Algorithm \ref{algo:Distributed-maxweight}, where every producer simply allocates its resource to the connected consumer with the largest demand.

%\IncMargin{0.2em}
\begin{algorithm}[t]
	\nl $t\in\{0,1,2,\ldots\};$ $\%$ Time-slotted algorithm.\\%
	\nl $\%$ Each producer serves her downstream consumers:\\%
	\nl \For{$j=1:N$}{
		\nl Find $i^*=arg\max_{i\in\mathcal{N}_j}X_i(t)$.\\%
		\nl Set $I_{ji^*}(t)=1$.\\%
	}
	\nl $\%$ Each consumer informs her neighbors for her demands:\\%
	\nl \For{$i=1:N$}{
		\nl $X_i(t)=[X_i(t-1)-M_i(t) ]^++A_i(t)$.\\%
		\nl Send $X_i(t)$ to every $j\in\mathcal{N}_i$.
	}	
	\caption{Max-weight Servicing Policy}\label{algo:Distributed-maxweight}
\end{algorithm}\DecMargin{1em}
%%%%%%%%%%%\vspace{-2mm}

An example of two cooperating EE and the benefits that emanate from their collaboration is depicted in Figure \ref{fig:queue-example-2}. We observe that cooperation does not increase the maximum aggregate request rate of the EE yet the set of supportable rates expands significantly due to the flexibility of re-routing the requests among the entities.

\section{Production Scheduling} \label{sec:production}

We now focus on the case with $\mathcal{K}=\{1,2,\ldots,K\}$ different commodities. Each economic entity $i$ generates $A_{ik}(t)$ requests for commodity $k$ during slot $t$, where the processes $A_{ik}$, $i\in\mathcal{N}$, $k\in\mathcal{K}$ are i.i.d. and independent, with $E[A_{ik}(t)]=a_{ik}\leq A_{max}$. Each EE $j\in\mathcal{N}$ has a certain set of feasible production plans $\mathcal{P}_j$. Under each plan $p\in\mathcal{P}_j$ the EE produces $B_{jk}^{p}(t)=B_{jk}^{p}\leq B_{max}$ units of commodity $k\in\mathcal{K}$ in each slot $t$. Although we consider deterministic production, our results can be directly extended for stochastic production where, for example, the schedule selects only the mean values. Finally, we assume that each entity can update her production plan every time period $t=nT$ with $T>>1$. This reflects practical system constraints where production scheduling cannot follow the dynamics of request generation and resource allocation.

%where $B_{jk}^{p},\,\forall j\in\mathcal{N}, k\in\mathcal{K}, p\in\mathcal{P}_j$ are assumed i.i.d. and independent processes with $E[B_{jk}^{p}(t)]=b_{jk}^{p}$. 

%% 17/4 We model this system as a network of queues where there is a queue $(i,k)$ for the requests generated at each consumer $i$ for commodity $k$; and a set of producers $(j,k)$ for every EE $j\in\mathcal{N}$ that serves commodity $k\in\mathcal{K}$. An example is shown in Fig. \ref{fig:queue-example-2-2}. 

The control policies of this economy include both the service allocation and the production planning. We define $Z_{jp}(nT)\in\{0,1\}$ as the decision of EE $j$ to select plan $p\in\mathcal{P}_j$ during period $nT$. This yields the production vector:
\begin{align} 
	\bm{B}_{j}^{p}=(B_{j1}^{p}, B_{j2}^{p},\ldots,B_{jK}^{p})\,.
\end{align}
We denote with $\bm{Z}$ the production vector of all EEs, and define the set of all feasible plans:
\begin{equation}
	\mathcal{Z}=\{\bm{Z}: Z_{jp}\in\{0,1\},\,\sum_{p\in\mathcal{P}_j}Z_{jp}\leq 1,\,j\in\mathcal{N}\,\}\,.
\end{equation}

It is assumed that each producer can satisfy only the demands of one consumer (including itself) for each commodity, but can concurrently serve more than one other consumers for different commodities. Therefore, the set of all feasible control policies is
\begin{equation}
	\mathcal{I}_{\mathcal{K}}=\big(\bm{I}_{\mathcal{K}}: I_{ji}^{k}\in\{0,1\},\,\,\,\sum_{i\in\mathcal{N}_j}I_{ji}^{k}\leq 1,\,\forall\,j\in\mathcal{N},k\in\mathcal{K} \big). \nonumber
\end{equation}
Under the above assumptions, the amount of resource $k$ that consumer $i$ receives during slot $t$ is:
\begin{equation}
	M_{ik}(t)=\sum_{j\in\mathcal{N}_i}I_{ji}^{k}(t)\sum_{p\in\mathcal{P}_j}Z_{jp}(t_T)B_{jk}^{p}(t),
\end{equation}
where $t_T=\left \lfloor{\frac{t}{T}}\right \rfloor $ is the last time before slot $t$ that the production schedule was updated. We denote $\bm{M}_{\mathcal{K}}(t)=(M_{ik}(t):i\in\mathcal{I},k\in\mathcal{K})$ the $I\times K$ matrix of services in slot $t$ which depends on the planning and allocation decisions $\bm{Z}$ and $\bm{I}_{\mathcal{K}}$ in that slot.

\begin{figure}[t]
	\centering
	\subfigure[]{\includegraphics[width=0.205\textwidth]{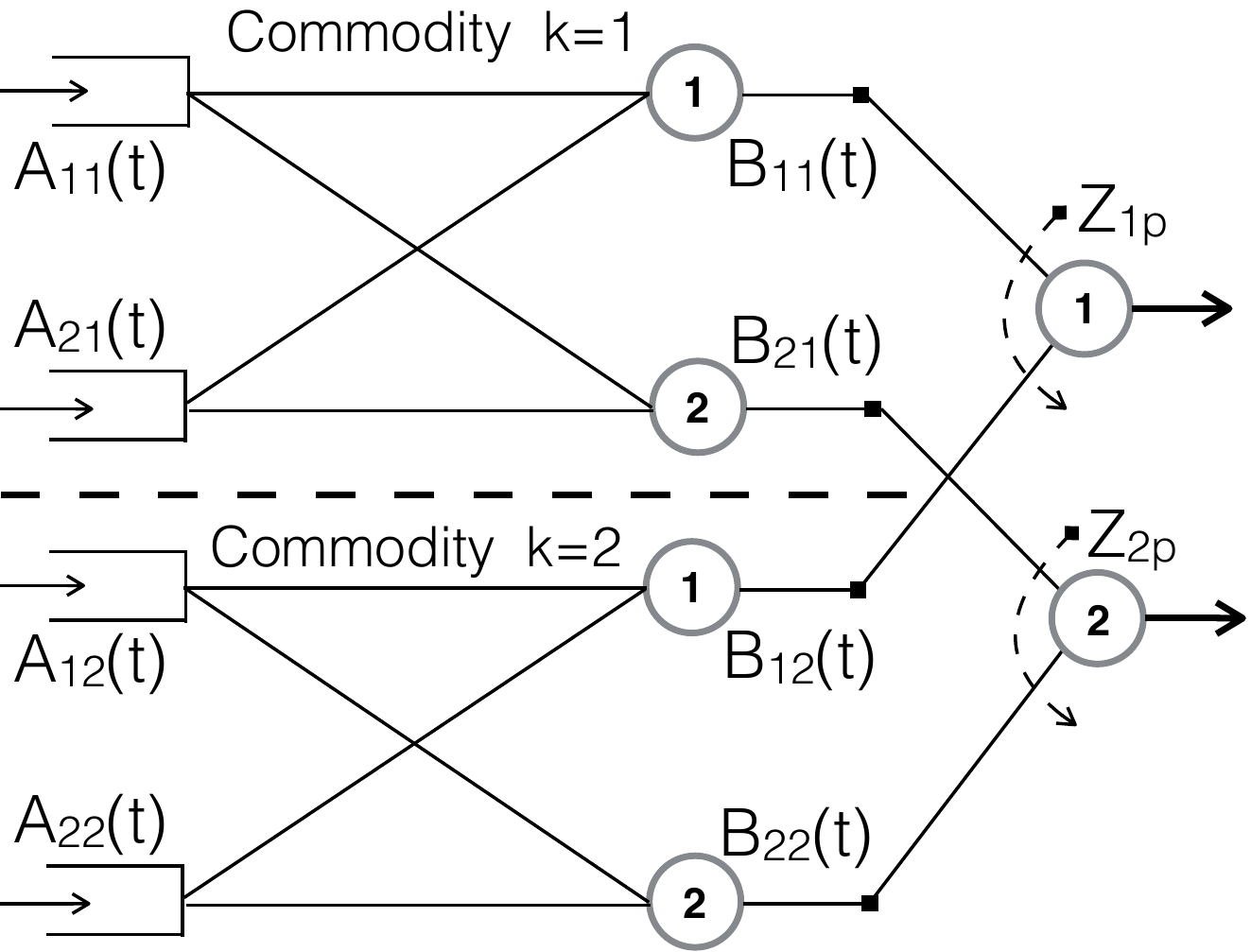}}
	\subfigure[]{\includegraphics[width=0.24\textwidth]{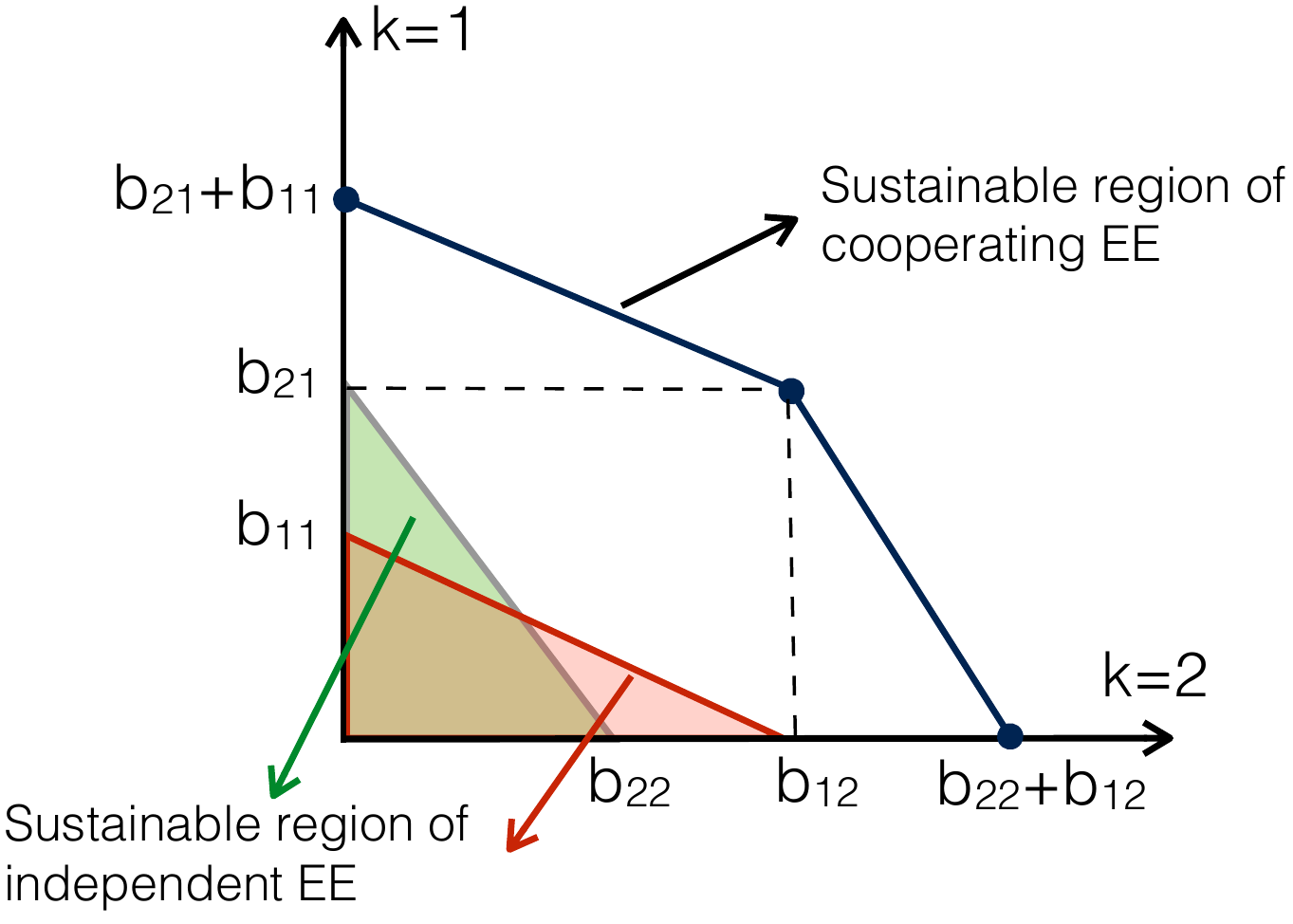}}
	\caption{\small{(a): Example of $K=2$ commodities for system in Fig. \ref{fig:queue-example-2}(a). The eligible plans for EE $j=2$ are $p_1=(b_{21}, b_{22}=0)$, $p_2=(b_{21}=0, b_{22})$ and similarly for $j=1$. (b): The sustainable demand region $\bm{\Lambda}_{\mathcal{K}}^{\mathcal{P}}$} of the system when EEs cooperate and when they operate independently. Each axis represents the total served amount for each commodity in both EE. }\label{fig:queue-example-2-2}
\end{figure}

The unsatisfied demands at each consumer for every commodity evolve in time as follows:
\begin{equation}
	X_{ik}(t)=[X_{ik}(t-1)-M_{ik}(t) ]^++A_{ik}(t)\,.
\end{equation}
We can therefore define the \emph{sustainability region} \cite{tassiulas-book}:
\begin{align}
	\bm{\Lambda}_{\mathcal{K}}^{\mathcal{P}}=&\{\bm{a}_{\mathcal{K}}=(a_{ik}: i\in\mathcal{N},k\in\mathcal{K}):\nonumber \\
	&\bm{a}_{\mathcal{K}}\in Co\{\bm{M}_{\mathcal{K}}(\bm{I}_{\mathcal{K}},\bm{Z})\,|\,\,
	\bm{I}\in\mathcal{I}_{\mathcal{K}},\bm{Z}\in\mathcal{Z}\}\}\,, \nonumber
\end{align}
where $Co(\cdot)$ is the convex hull operator. The following lemma holds.
%\IncMargin{0.2em}
\begin{algorithm}[t]
	\nl $t=nT,\,n\in\{0,1,2,\ldots\};$ $\%$ Time periods.\\%
	\nl $\%$ Each producer selects her production plan:\\%
	\nl \For{$j=1:N$}{
		\nl $\%$ Find highest commodity $k^*$ demand $\forall i\in\mathcal{N}_j$.\\%		
		\nl Set $\hat{X}_{ik^*}(t)=X_{ik^*}(t)$; else $\hat{X}_{ik}(t)=0$ \\%
		\nl Find $p^*=arg\max_{p\in\mathcal{P}_j}\sum_{i\in\mathcal{N}_j}\sum_{k\in\mathcal{K}}\hat{X}_{ik^*}(t)B_{jk}^{p}$.\\%
	}
	\nl $\%$ Each producer allocates her resources in each slot $\tau$:\\%
	\nl \For {$\tau=t:t+T-1$}{
		\nl \For{$j=1:N$}{ 
			\nl \For{$k=1:K$}{
				\nl Find $i^*=arg\max_{i\in\mathcal{N}_j}X_{ik}(\tau)B_{jk}^{p^*}$.\\%
				\nl Set $I_{ji^*}^{k}(\tau)=1$.\\%
			}	
		}
		\nl $\%$ Each consumer updates its pending demands:\\%
		\nl \For{$i=1:N$}{
			\nl \For{$k=1:K$}{
				\nl $X_{ik}(\tau)=[X_{ik}(\tau-1)-M_{ik}(\tau) ]^++A_{ik}(\tau)$.\\%
				\nl Send $X_{ik}(\tau)$ to every $j\in\mathcal{N}_i$.
			}
		}
	}		
	\caption{Production and Service Allocation Policy}\label{algo:Distributed-production}
\end{algorithm}%\DecMargin{1em}

\begin{lemma}
	For any demand generation matrix $\bm{a}_{\mathcal{K}}\in\bm{\Lambda}_{\mathcal{K}}^{\mathcal{P}}$, there exists a stationary randomized control policy $\Pi_{opt}^{K}$ that chooses $\bm{Z}$ every $T$ slots, and $\bm{I}_{\mathcal{K}}$ every slot $t$, and stabilizes the economy.
\end{lemma}

Next, we design a dynamic policy that determines the plan and the service allocation, and stabilizes the system whenever $\bm{a}\in\bm{\Lambda}_{\mathcal{K}}^{\mathcal{P}}$. This is non-trivial since it involves decisions in different time-scales. Algorithm \ref{algo:Distributed-production} describes the policy. The main idea is that in every time period, each producer finds for each of her neighboring consumers the commodity with the largest pending requests (line 5), and then uses these to select the plan that will serve better the neediest consumers (line 6). Then, in every time slot, the producers allocate their resources to the largest backlog of demands (lines 10-12). Note that the algorithm can be executed in a distributed (but synchronous) fashion. The following theorem states its performance.
\begin{theorem}
	Suppose an economy has a sustainability region $\bm{\Lambda}_{\mathcal{K}}^{\mathcal{P}}$ and demand $\bm{a}_{\mathcal{K}}$ such that $\bm{a}_{\mathcal{K}}+\epsilon\bm{1}\in\bm{\Lambda}_{\mathcal{K}}^{\mathcal{P}}$; then under the policy described in Algorithm	\ref{algo:Distributed-production} and a set value for $V$, the economy is sustainable and the average backlog of unsatisfied demands is bounded:
	\begin{equation}
		\lim_{t\rightarrow \infty}\frac{1}{t}\sum_{\tau=0}^{t-1}\sum_{i,k}E[X_{ik}(\tau)]\leq \frac{NKA_{max}^{2}+\sum_{i,k}\sum_{j\in\mathcal{N}_i}\big(B_{jk}^{p*}\big)^{2}}{\epsilon/T} \nonumber
	\end{equation}
	where $B_{jk}^{p^*}$ is the max production rate of $j$ for commodity $k$.
\end{theorem}
\noindent The backlog of unsatisfied demands increases with $T$ and $K$; and depends on the exchange graph as it includes the in-degree of nodes.

\section{Costly Production Schedules} \label{sec:production-costs}

We now extend our analysis to the case of costly production plans. Clearly, when production of resources induces significant costs, then it is not ensured that demand rates within the sustainability region of the system will be served (as in the previous section). Namely, when self-interested EEs cooperate it is expected that each one of them will attempt to satisfy her demands through this collaboration (e.g., by receiving resources from others) while at the same time minimizing her own production cost (e.g., selecting low-cost plans). It is well known that such free-riding behaviors may lead to tragedy of commons phenomena, where the cooperation benefits diminish rapidly. 

To address this issue, we introduce an additional design criterion for our cooperation policy, namely that of being incentive-compatible (IC) and hence implementable by the self-interested EEs. However, it is well-known that such game theoretic-based policies require the a priori knowledge of the demand rates and the cooperation benefits, information which in many practical settings is not available. We present here a production and service exchange algorithm that ensures the \emph{sustainable and incentive compatible} operation of the economy, whenever that is possible, without relying on prior information about demands (or, production rates).

In detail, we assume that when an EE $j\in\mathcal{N}$ selects plan $p\in\mathcal{P}_j$, she incurs cost of $c_{jp}$ units for that period. Therefore, when $j$ operates independently, her optimal production policy can be found by solving the optimization problem $(Opt_{j}^{ind})$:
\begin{equation}
	\min_{\{\zeta_{jp}\} } \sum_{p\in\mathcal{P}_j}c_{jp}\zeta_{jp}\,\,\,\,
	s.t.\,\,\,\, a_{jk}\leq\sum_{p\in\mathcal{P}_j}\zeta_{jp}B_{jk}^p,\,\forall k\in\mathcal{K}, \label{eq:independent-performance}
\end{equation}
where $\zeta_{jp}\in[0,1]$ is the probability to select plan $p\in\mathcal{P}_j$. We denote the min-cost planning solution $J_{j}^{ind}(\bm{a}_j)$, where $\bm{a}_j=(a_{jk}:k\in\mathcal{K})$. 

When the entities collaborate they select their production and servicing policies in an incentive-compatible fashion that also ensures sustainability of the economy. In order to satisfy the IC requirement we leverage the Nash bargaining solution (NBS) \cite{bargaining} which possesses the important properties of axiomatic fairness and Pareto efficiency (see \cite{mazumdar} for a discussion), and hence is self-enforcing in the presence of strategic entities. The solution of following optimization problem $(Opt_{N})$ describes the NBS solution when the resources and demands are known; it will serve as a benchmark for assessing the performance of our dynamic policy in the sequel:

\begin{equation}
	\max_{ \{\rho_{ji}^k\}, \{\zeta_{jp}\} } H=\Pi_{j=1}^{N}\big(J_{j}^{ind}(\bm{a}_j)- \sum_{p\in\mathcal{P}_j}c_{jp}\zeta_{jp} \big) \label{eq:static-objective}
\end{equation}
s.t.
\begin{align}
	&a_{ik}+\epsilon_1 \leq \sum_{j\in\mathcal{N}_i}\rho_{ji}^k\sum_{p\in\mathcal{P}_j}\zeta_{jp}B_{jk}^p,\,\,\forall\,\,i\in\mathcal{N},\,k\in\mathcal{K}\,,\\
	&\sum_{i\in\mathcal{N}_j}\rho_{ji}^{k}\leq 1,\,\,\forall j\in\mathcal{N}\,,k\in\mathcal{K}\,,\\
	&\sum_{p\in\mathcal{P}_j}c_{jp}\zeta_{jp}+\epsilon_2\leq J_{j}^{ind}(\bm{a}_j),\,\,\,\forall\,j\in\mathcal{N}\,, \\
	&\sum_{p\in\mathcal{P}_j}\zeta_{jp}\leq 1,\,\,\,\forall\,j\in\mathcal{N}\,, \\
	&\rho_{ji}^k,\,\zeta_{jp}\in\,[0,1],\,\,\forall\,i,j\in\mathcal{N},k\in\mathcal{K}\,.
\end{align}
\vspace{0.5mm}
$\rho_{ji}^k$ is the probability that $j$ will serve the demands of $i$ for commodity $k$; and $\zeta_{jp}$ the probability she will select plan $p$. We use constants $\epsilon_1, \epsilon_2>0$ to avoid limiting cases. This problem admits a solution as we assume zero cooperating costs (e.g., no need for additional infrastructure). We denote with $\Lambda_{\mathcal{K}}^{\mathcal{P},B}$ the closure of set of demand rates that can be served under the NBS (solution of $P_{S}$) with bounded costs. In the case $c_{jp}$ parameters are constant and finite, this region coincides with $\Lambda_{\mathcal{K}}^{\mathcal{P}}$. It is easy to show (similarly to Lemma 1 and 2) that the policies stemming from $(Opt_{N})$ are necessary, sufficient and optimal for the incentive-compatible and sustainable operation of the economy; and we denote $H_{N}^{*}(\bm{a}_{\mathcal{K}})$ the solution of $(Opt_{N})$.

Our goal is to design a dynamic policy that asymptotically approaches arbitrary close to the optimal planning and servicing solution. Algorithm \ref{algo:cost-production} describes the dynamic policy. First, note that we introduce an auxiliary variable $Y_{j}(t)$, for each EE $j$, representing the evolution of a virtual queue \cite{neely-energy} which, when stable, ensures the following constraint:
\begin{equation}
	\lim_{t\rightarrow \infty}\frac{1}{t}\sum_{\tau}E\{ \sum_{p\in\mathcal{P}_j}c_{jp}Z_{jp}(\tau)\}\leq \lim_{t\rightarrow \infty}\frac{1}{t}\sum_{\tau}E\{J_{j,\tau}^{ind}\}, 
\end{equation}
where $J_{j,\tau}^{ind}$ is the cost every EE $j$ would incur in each period if she was serving only her own needs. This quantity is not known in advance, but can be computed by the entity in every period after observing the pending and newly generated demands. This means that each EE has to run an algorithm for solving problem $(Opt_{j}^{ind})$, e.g., using a threshold-based decision policy (see \cite{tassiulas-book} for examples), and use the calculated value in Algorithm 3 where $\overline{J_{j,t}^{ind}}$ is the respective running average. The virtual queue evolves in successive periods:
\begin{equation}
	Y_{j}((n+1)T)=[ Y_{j}(nT)-J_{j,nT}^{ind} ]^+ +\sum_{p\in\mathcal{P}_j}c_{jp}Z_{jp}(nT) \,.
\end{equation}
As before, each producer finds the most demanding commodity for each of her neighbors (line 5) and then selects the plan that maximize the expression (\ref{eq:alg3-expression}): namely, the plan is selected so as to maximize the offered service (last term), and minimize the cost of production (second term), while balancing the cost under cooperation with the cost the EE would incur if operating independently. Next, each EE allocates the service, in each small slot, based on the accumulated demands of her neighbors (lines 8-12), for each commodity, and informs her neighbors about the updated pending demands (lines 13-17). At the end of the period it updates the virtual queue (line 20) in order to ensure (asymptotically) that the cooperation cost will be bounded by the independent cost.

The next theorem characterizes the algorithm's performance in terms of cost and average backlog demands.
\begin{algorithm}[t]
	\nl $t=nT,\,n\in\{0,1,2,\ldots\};$ $\%$ Time periods.\\%
	\nl $\%$ Each producer selects her production plan:\\%
	\nl \For{$j=1:N$}{
		\nl $\%$ Find highest commodity $k^*$ demand $\forall i\in\mathcal{N}_j$.\\%		
		\nl Set $\hat{X}_{ik^*}(t)=X_{ik^*}(t)$; else $\hat{X}_{ik}(t)=0$ \\%
		\nl Set $Z_{jp^*}(t)=1$ for $p^*$ that maximizes:
		\begin{align}
			&V\big(\overline{J_{j,t}^{ind}}-\sum_{p\in\mathcal{P}_j}c_{jp}Z_{jp}(t)\big) -\nonumber \\ 
			&2\sum_{p\in\mathcal{P}_j}Z_{jp}\big(Y_j(t)c_{jp}-T\sum_{k=1}^{K}\sum_{i\in\mathcal{N}_j}\hat{X}_{ik}(t)B_{jk}^p\big)  \label{eq:alg3-expression}
		\end{align}
	}
	\nl $\%$ Each producer allocates her resources in each slot $\tau$:\\%
	\nl \For {$\tau=t:t+T-1$}{
		\nl \For{$j=1:N$}{ 
			\nl \For{$k=1:K$}{
				\nl Find $i^*=arg\max_{i\in\mathcal{N}_j}X_{ik}(\tau)B_{jk}^{p^*}$.\\%
				\nl Set $I_{ji^*}^{k}(\tau)=1$.\\%
			}	
		}
		\nl $\%$ Each consumer updates its demands:\\%
		\nl \For{$i=1:N$}{
			\nl \For{$k=1:K$}{
				\nl $X_{ik}(\tau)=[X_{ik}(\tau-1)-M_{ik}(\tau) ]^++A_{ik}(\tau)$.\\%
				\nl Send $X_{ik}(\tau)$ to every $j\in\mathcal{N}_i$.
			}
		}
	}	
	\nl \For{$j=1:N$}{
		\nl $\%$ Each consumer updates its cost queue:\\%
		\nl $Y_{j}(t)=[Y_{j}(t)-J_{j,t}^{ind}(t)]^++\sum_{p\in\mathcal{P}_j}c_{jp}Z_{jp}(t)$.\\%	
	}
	\caption{Policies for costly production plans}\label{algo:cost-production}
\end{algorithm}\DecMargin{1em}

\begin{theorem}
	Suppose an economy has a sustainability and IC region $\bm{\Lambda}_{\mathcal{K}}^{\mathcal{P},B}$ and demand $\bm{a}_{\mathcal{K}}$ such that $\bm{a}_{\mathcal{K}}+\epsilon\bm{1}\in\bm{\Lambda}_{\mathcal{K}}^{\mathcal{P},B}$; then under the policy described in Algorith	\ref{algo:cost-production} the economy is sustainable and the cooperation policy is incentive compatible, with optimality and demand backlogs bounded as follows:
	\begin{equation}
		\lim_{t\rightarrow \infty}\frac{1}{t}\sum_{\tau=0}^{t-1}\sum_{i,k}E[X_{ik}(\tau)]\leq
		\frac{C+VG_{max}}{\epsilon(\bm{a}_{\mathcal{K}})} 
	\end{equation}
	\begin{equation}
		E\{H(t)\}\geq H_{N}^{*}(\bm{a}_{\mathcal{K}}) -\frac{C}{V}
	\end{equation}
	where $C= TKNA_{max}^{2}+TK\sum_{i}\big(d_{i}^{in}B_{max}^2+2(c_{ip}^{max})^2\big)$; $G_{max}$ is the maximum value of eq. (\ref{eq:static-objective}), i.e., the product of the highest cost values $c_{ip}^{max}$ of all entities.
\end{theorem}

Interestingly, with Algorithm 3 that is amenable to distributed execution, the economic entities achieve a performance that is arbitrary close to the objective of $Opt_{N}$, while the average number of pending demands is bounded. Therefore, the EEs ensure the sustainability of the economy and enforce an $\epsilon_0$-NBS solution, where $\epsilon_0=C/V$. The balance between fairness (or incentive compatibility) and backlog can be determined by proper selection of $V$. Also, it is interesting to note that our policy uses the benchmark independent performance which does not have to be known in advance, but it suffices to use its running average (which eventually will converge in the expected value).

\section{Discussion and Conclusions} \label{sec:conclusions}

The problem of cooperation lies at the core of our social and economic life, and has been excessively studied with a recent focus on the impact of the network graph on the bargaining power of each entity \cite{bonacich} or the cooperation outcome \cite{rand-coop}. Similarly, from an engineering point of view, there is an interesting literature proposing cooperation models and analyzing their equilibriums, e.g., for decentralized sharing of wireless services \cite{guerin-upns} or infrastructure sharing \cite{courcoubetis}. Two important and particularly challenging aspects that remain to be understood are (i) the dynamics of cooperation, namely how such equilibriums can be achieved in an on-line fashion; and (ii) the impact of network or graph constraints on the performance of such cooperative schemes. 

Motivated by these observations, our work proposes a Lyapunov-based optimization approach for designing cooperation policies that achieve asymptotically an efficient and incentive compatible equilibrium. Importantly, this solution handles well information uncertainties as it does not presume the existence of prior information about the needs of the entities, nor the benefits that their collaboration can achieve. The advantages of our approach come at the expense of asymptotic optimality which, moreover, in the case of costly production plans achieves a near-optimal outcome - and hence close to the bargaining equilibrium. Another important point here is the time-scale separation we considered, where we followed an analysis similar to \cite{neely-infocom12}, \cite{giannakis-2scale}. The impact on performance can be directly seen at the respective demand bounds, and this calls for further research in order to improve that result.

The problem of exchanging a single commodity in a static environment was studied in our previous work \cite{iosifidis-sigmetrics}, where the existence of competitive and coalitional equilibriums was proved. From a different perspective, the benefits of cooperation and resource pooling in servicing systems have been studied in operations research problems. Often the question there is how the induced cost will be split among the entities, e.g., see \cite{anily}, while more recent studies \cite{murrel} argue that pooling might even reduce the overall performance under some assumptions about the dependence of the servicing costs on the total load. Unlike our approach, these important works focus on static systems with known demands and capacities. Besides, we consider the selection of production plans and multiple commodities. This is a particularly important aspect as it reveals that diversity in production is particularly beneficial for cooperative systems, an argument that is both intuitive and experimentally validated in macroscopic scale \cite{hidalgo-pnas}.

\section{Acknowledgments}

This research was sponsored by the U.S. Army Research Laboratory and the U.K. Ministry of Defence under Agreement Number W911NF-16-3-0001. The views and conclusions contained in this document are those of the authors and should not be interpreted as representing the official policies, either expressed or implied, of the U.S. Army Research Laboratory, the U.S. Government, the U.K. Ministry of Defence or the U.K. Government. The U.S. and U.K. Governments are authorized to reproduce and distribute reprints for Government purposes notwithstanding any copy-right notation hereon. Finally, the authors acknowledge the by the National Science Foundation under Grant CNS 1527090.

\section*{APPENDIX}

\noindent \underline{\textbf{Proof of Lemma 1}}: The necessity of conditions (\ref{eq:stability-condition}) is straightforward. For the sufficiency part, we need to show that whenever these conditions are satisfied, we can find a randomized policy that supports $\bm{a}$. Let $\bm{\rho}=(\rho_{ji}:\,(j,i)\in\mathcal{E})$ denote the randomized policy where $\rho_{ji}\geq0$ is the probability producer $j$ will serve consumer $i$. It is then: 
\begin{equation}
	\sum_{i\in\mathcal{Q}}(a_i+\epsilon)\leq \sum_{j\in\mathcal{N}_\mathcal{Q}}\rho_{ji}b_j,\,\,\,\forall \mathcal{Q}\subseteq\mathcal{N},\,\,\,
	\sum_{i\in\mathcal{N}_j}\rho_{ji}\leq 1,\,\,\,\forall\,i,j\in\mathcal{N}\nonumber
\end{equation}
Finding a randomized policy that satisfies the above constraints, and for the minimum possible value of $\epsilon$ is a linear program. Moreover, for the specific example here, we can show the existence of the randomized policy through construction.

In particular, we construct a network and employ the max-flow/min-cut theorem. The graph has a source and sink node $S$ and $D$. From the $S$ we have $i=1,\ldots,N$ links, with capacity $a_1,a_2,\ldots,a_N$, respectively, connecting it with the 2nd-layer nodes. Each of the latter nodes is connected to a subset of $N$ nodes at the 3rd layer (having capacity $a_i$, respectively), based on the links $\mathcal{E}$ in the original graph $G$. Finally, the 3rd-layer nodes are connected to the $D$ with links of capacity $b_1,b_2,\ldots,b_N$. The minimum cut of this graph is comprised of the outgoing edges of node $S$, with total capacity $\sum_{i=1}^{N}a_i$. Hence, the maximum network flow can support the given vector $\bm{a}$. In particular, let $f_{ij}$ denote the flow over link $(i,j)$, and $f_1,f_2,\ldots,f_N$ the flows emanating from $S$, under a max-flow solution. Since the graph satisfies (\ref{eq:stability-condition}) the max-flow is supportable, and the routing policy for each link can be simply defined as the ratio $f_{ij}/a_i$. The max-flow solution can be found in polynomial time. Q.E.D.

\vspace{2mm}
\noindent \underline{\textbf{Proof of Theorem 1}}: We use the Lyapunov function,
\begin{equation}
	L(\bm{X}(t))=\frac{1}{2}\sum_{i=1}^{N}\big(X_i(t)\big)^2 \nonumber
\end{equation}
and following the analysis in \cite{tassiulas-book} we define the Lyapunov drift $\Delta\big(L(\bm{X}(t))\big)$ for which it holds:
\begin{align}
	&\Delta\big(L(\bm{X}(t))\big)=E\{L(\bm{X}(t+1))-L(\bm{X}(t))|\bm{X}(t)\}\leq\nonumber \\
	&\sum_{i=1}^{N}\frac{A_{max}^2 + M_{i,max}^2}{2}+\sum_{i=1}^{N}a_iX_i(t)  - E\{\sum_{i=1}^{N}X_i(t)M_i(t)|\bm{X}(t) \}\,, \nonumber
\end{align}
where $M_{i,max}$ is the maximum service node $i$ can receive, namely $M_{i,max}=d_{i}^{in}B_{max}$, where $d_{i}^{in}$ is the in-degree of $i$.

Hence, selecting in each slot the servicing policy that maximizes the last term, ensures a negative Lyapunov drift which in turn (see Lemma 4.1 \cite{tassiulas-book}) ensures strong stability and a long-term expected backlog:
\begin{equation}
	\lim_{t\rightarrow \infty}\frac{1}{t}\sum_{\tau=0}^{t-1}\sum_{i=1}^{N}E[X_i(\tau)]\leq \frac{NA_{max}^{2}+\sum_{i=1}^{N}d_{i}^{in}B_{max}^{2}}{2\epsilon(\bm{a})} \nonumber
\end{equation}	
and hence that the cooperative network is sustainable.

\begin{figure}
	\centering
	\includegraphics[width=0.31\textwidth]{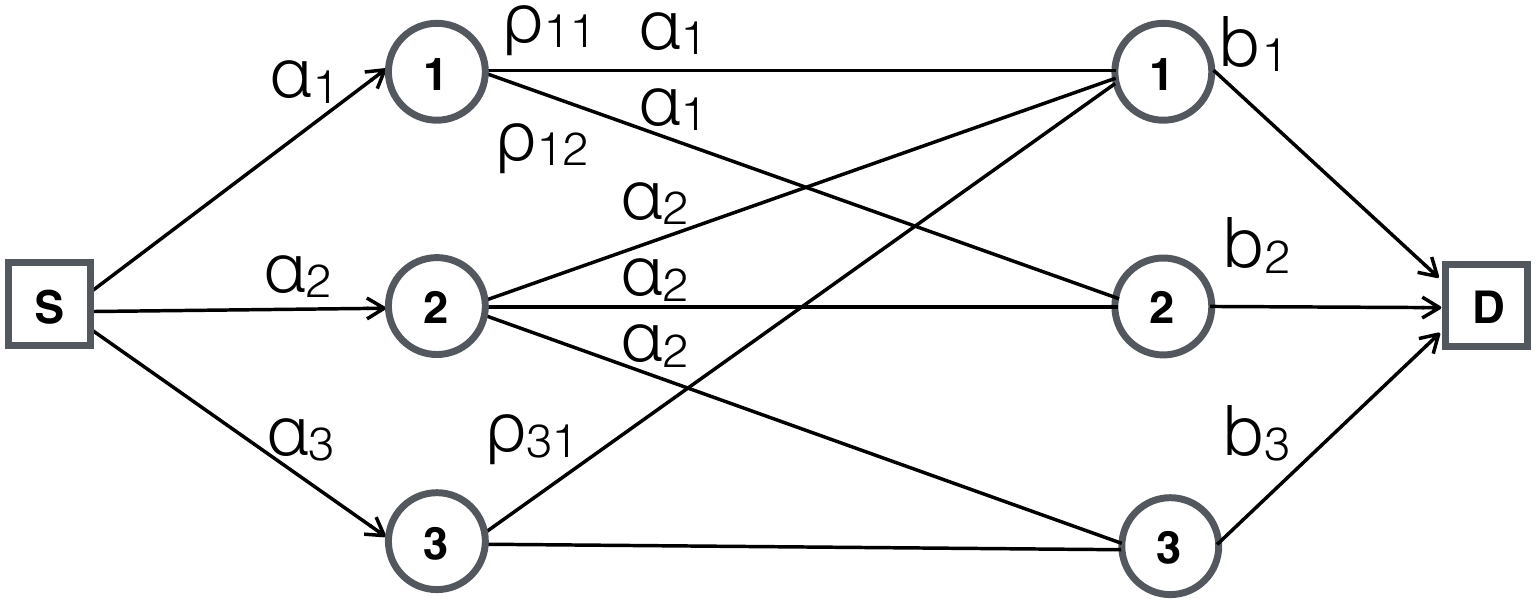}
	\caption{\small{The max-flow diagram for the economy of Fig. \ref{fig:queue-example}.  }}\label{fig:max-flow-1-commodity}
\end{figure}

Let us now discuss why this policy can be devised in a distributed fashion. This boils down to whether quantity:
\begin{equation}
	\sum_{i=1}^{N}X_i(t)\sum_{j\in\mathcal{N}_i}I_{ji}(t)B_{j}(t)\,,
\end{equation}
can be maximized in each slot $t$ distributively by the nodes. In each slot $t$, every producer $j$ can observe her available resources $B_{j}(t)$ and the demands of her connected consumers $X_i(t),\,i\in\mathcal{N}_j$, and serve the one with the largest backlog. Since the allocation decisions of the producers are independent, this will lead to serving the consumers with the largest pending requests in each time instance. Hence, the proposed policy will asymptotically achieve the goal of negative drift. Clearly, there will likely be slots where excessive resource will be allocated to a certain consumer, e.g., when a producer has more available resources than those needed, or when two or more producers will concurrently serve the same consumer. Q.E.D.

\vspace{2mm}
\noindent \underline{\textbf{Proof of Lemma 2}}: It is easy to show both the necessity and the sufficiency of conditions in the Lemma. For the sufficiency part, we need to show that whenever these conditions are satisfied, we can find a randomized policy that supports the admissible rate vector $\bm{a}_{\mathcal{K}}$. This can be proved using Caratheodory's theorem and observing that we can construct a state-independent stochastic planning and service allocation policy that implements $\Pi_{\mathcal{K}}$. Q.E.D.

\vspace{2mm}
\noindent \underline{\textbf{Proof of Theorem 2}}: We use again a quadratic Lyapunov function and work with the $T$-slot drift:
\begin{align}
	\Delta_{T}(t)=E\{L\big(\bm{X}_{\mathcal{K}}(t+T)\big)-L\big(\bm{X}_{\mathcal{K}}(t)\big)|\bm{X}_{\mathcal{K}}(t)\}\,,
\end{align}
where we observe the evolution of demands at each consumer for every commodity over $T$ slots beyond time instance $t$. Using the following result (Lemma 4.3 \cite{tassiulas-book}):
\begin{equation}
	V\leq \max[U-\mu,0]+A \Rightarrow V^2\leq U^2+\mu^2+A^2-2U(\mu-A), \nonumber
\end{equation}
and summing over all slots $\tau$ within period $T$, we find that:
\begin{align}
	&X_{ik}(t+T-1)^2 -X_{ik}(t)^2\leq \sum_{\tau=t}^{t+T-1}M_{ik}^2(\tau)+ \sum_{\tau=t}^{t+T-1}A_{ik}^2(\tau)\nonumber \\ 
	&-2\sum_{\tau=t}^{t+T-1}X_{ik}(\tau)[M_{ik}(\tau)-A_{ik}(\tau)] \nonumber .
\end{align}
Hence, the drift is bounded as follows
\begin{align}
	&\Delta_{T}(t) \leq T\sum_{i,k} A_{max}^2 + T\sum_{i,k} M_{ik,max}^2 \nonumber \\ &-2E\{\sum_{i,k}\sum_{\tau=t}^{t+T}X_{ik}(\tau)[M_{ik}(\tau)-A_{ik}(\tau)]|\bm{X}_{\mathcal{K}}(t) \} \label{eq:drift-T}
\end{align}
where $M_{ik,max}$ is the maximum amount of resources of type $k$ that can be allocated to consumer $i$ in each slot, and can be written 
\begin{equation}
	M_{ik,max}=d_{i}^{in}B_{max},\, \text{or},\, M_{ik,max}=\sum_{j\in\mathcal{N}_{i}}\max_{p}\{B_{jk}^{p}\}\,.
\end{equation} 
Note that the first two terms in (\ref{eq:drift-T}) are upper bounded, and that the expression includes the values of matrix $\bm{X}_{\mathcal{K}}(\tau)$, $\tau=t+1, ...t+T$ which, clearly, are not available at the beginning of the period, i.e., $t=T$. Following the approach in \cite{neely-infocom12}, \cite{giannakis-2scale} we use for the entire period the values of pending demands at $t$, and hence we get:
\begin{align}
	&\Delta_{T}(t) \leq T\sum_{i,k} A_{max}^2 + T\sum_{i,k} M_{ik,max}^2 \nonumber \\
	&+2T\sum_{i,k}X_{ik}(t)a_{ik}-2E\{\sum_{i,k}X_{ik}(t)\sum_{\tau=t}^{t+T}M_{ik}(\tau)|\bm{X}_{\mathcal{K}}(t) \}\,, \nonumber 
\end{align}
and the latter term can be further written as:
\begin{align}
	E\{\sum_{i,k}X_{ik}(t)\sum_{j\in\mathcal{N}_i}\sum_{\tau=t}^{t+T-1}I_{ji}^{k}(\tau)\sum_{p\in\mathcal{P}_j}Z_{jp}(t)B_{jk}^{p}|\bm{X}_{\mathcal{K}}(t)\}\,.
\end{align}
The goal of our policy is to maximize this quantity opportunistically, where production planning decisions are made every period $t=nT,n=0,1,\ldots$, and service allocation every slot $t=nT+\tau$. Since the demand generation rates are within the sustainability region, it is easy to see that the result follows (see Lemma 4.1 in \cite{tassiulas-book}). Q.E.D.

\vspace{2mm}
\noindent \underline{\textbf{Proof of Theorem 3}}: There are two different queues evolving as follows:
\begin{equation}
	X_{ik}(t+1)=[ X_{ik}(t)-M_{ik}(t)]^+ + A_{ik}(t),\,\,\forall\, (i,k)
\end{equation}
\begin{equation}
	Y_{i}(t+1)=[ Y_{i}(t)-J_{i,t}^{ind}]^+ +\sum_{p\in\mathcal{P}_i}c_{ip}Z_{ip}(t,\,\,\forall\, i\,.
\end{equation}
We define the following Lyapunov function:
\begin{equation}
	L(\bm{Y},\bm{X})=\frac{1}{2}\sum_{i,k}X_{ik}^{2}+\frac{1}{2}\sum_{i}Y_{i}^2\,.
\end{equation}
We consider the T-slot Lyapunov drift. Let us first observe that the evolution of the respective queues, can be bounded as follows:
\begin{align}
	&X_{ik}(t+T-1)^2-X_{ik}(t)^2 \leq \sum_{\tau=t}^{t+T-1}M_{ik}^{2}(\tau) +\nonumber \\ &\sum_{\tau=t}^{t+T-1}A_{ik}^{2}(\tau)-2\sum_{\tau=t}^{t+T-1}X_{ik}(\tau)[M_{ik}(\tau)-A_{ik}(\tau)]
\end{align}
and
\begin{align}
	& Y_i(t+T-1)^2-Y_i(t)^2\leq \big( J_{i,t}^{ind} \big)^2 + \nonumber \\
	& \big(\sum_{p\in\mathcal{P}_i}c_{ip}Z_{ip}(t) \big)^2 -2Y_{i}(t)[ J_{i,t}^{ind} - \sum_{p\in\mathcal{P}_i}c_{ip}Z_{ip}(t) ]\,.	
\end{align}
If we add the above inequalities and rearrange terms, we get for the right hand side:
\begin{align}	
	&\leq B_{ik} + 2\sum_{\tau=t}^{t+T-1}X_{ik}(\tau)A_{ik}-2Y_i(t)J_{i,t}^{ind}\nonumber \\
	&-2\sum_{\tau=t}^{t+T-1}X_{ik}(\tau)M_{ik}(\tau)+2Y_{i}(t)\sum_{p\in\mathcal{P}_i}c_{ip}Z_{ip}(t)\,,
\end{align}
where the constant $B_{ik}$ is:
\begin{align}
	B_{ik}=T(M_{ik,max})^2+T(A_{max})^2+(J_{i,max}^{ind})^2+(c_{ip,max})^2\,.\nonumber
\end{align}
We proceed by relaxing the time-slot dependency on $X_{ik}$ and on service allocation decisions $I_{ji}^{k}$ as in \cite{neely-infocom12}, \cite{giannakis-2scale}. In practice, this means we consider the suboptimal case where we decide these policies with the information we have at the beginning of the period. Clearly, the performance of the Algorithms is superior (and hence stable) as they adapt on a per-slot basis. Therefore, we get:
\begin{align}	
	&Y_i(t+T-1)^2+X_{ik}(t+T-1)^2-Y_i(t)^2-X_{ik}(t)^2 \leq \nonumber \\
	&B_{ik}+2TX_{ik}(t)A_{ik}(t)-2Y_i(t)J_{i,t}^{ind} \nonumber \\
	&-2TX_{ik}(t)M_{ik}(t)+2Y_{i}(t)\sum_{p\in\mathcal{P}_i}c_{ip}Z_{ip}(t)\,.
\end{align}
Hence, the T-slot Lyapunov drift $\Delta_{T}(t)$ is defined as:
\begin{align}
	E\{L\big(\bm{X}(t+T),\bm{Y}(t+T)\big) - L\big(\bm{X}(t),\bm{Y}(t)\big)| ( \bm{X}(t),\bm{Y}(t) ) \} \nonumber
\end{align}
Therefore, if we add also the objective of maximizing the bargaining product, we get the following drift-plus-penalty expression:
\begin{align}
	&\Delta_{T}(t)-VE\{ \Pi_{i=1}^{N} \big( J_{i,t}^{ind}-\sum_{p\in\mathcal{P}_i}c_{ip}Z_{ip}(t)\big)|\bm{X}(t),\bm{Y}(t) \} \leq \nonumber \\
	& B+2TE\{ \sum_{i,k}X_{ik}(t)A_{ik}(t)|\bm{X}(t),\bm{Y}(t) \} \nonumber \\
	&-2E\{ \sum_{i}Y_{i}(t)J_{i,t}^{ind}|\bm{X}(t),\bm{Y}(t)\} \nonumber \\
	& -2TE\{ \sum_{i,k}X_{ik}(t)M_{ik}(t)||\bm{X}(t),\bm{Y}(t) \} \nonumber \\
	&+2E\{ \sum_{i}Y_{i}(t)\sum_{p}c_{ip}Z_{ip}(t)|\bm{X}(t),\bm{Y}(t)\} \nonumber \\
	&-VE\{ \Pi_{i=1}^{N} \big( J_{i,t}^{ind}-\sum_{p\in\mathcal{P}_i}c_{ip}Z_{ip}(t)\big)|\bm{X}(t),\bm{Y}(t) \}\,.
\end{align}
It is clear from the above, following the analysis in \cite{tassiulas-book} (see Theorem 5.8) that our algorithm maximizes opportunistically the drift in each slot. Hence, the above quantity can be bounded by the respective solution of the static problem $(Opt_{N})$.

Some other important points for this theorem are the following. The impact of fixing the values of $X_{ik}(t)$ and $M_{ik}(t)$ in the beginning of the time period is not explicitly studied due to lack of space, but can be analyzed as in \cite{neely-infocom12}, \cite{giannakis-2scale}. However, it is proved that since the suboptimal policy (of deciding on a per period basis) stabilizes the system, the policy that updates decisions per slot also makes the economy sustainable, and possibly achieves a closer to optimal point.

Another interesting point is that each entity has to run in parallel a threshold-based algorithm in order to assess the cost $J_{i,\tau}^{ind}$ that would incur, had she operated in an independent mode. Then, using this quantity, it updates the running average:
\begin{equation}
	J_{i,t}^{ind}=\frac{1}{t}\sum_{\tau=0}^{t-1}J_{i,\tau}^{ind}\,,
\end{equation}
and plugs it in Algorithm 3. It is easy to see that, due to our assumptions about the demand generation processes, after some iterations this quantity will converge to the steady-state and a constant value. Finally, the Nash bargaining objective is a product of the terms for each entity. Since the decisions and cost functions of the EEs are decoupled, this product can be maximized in a decentralized fashion. Alternatively, one can employ the equivalent logarithmic formulation of the NBS introduced in \cite{mazumdar}. Q.E.D.

\end{document}